\documentclass[aps,prl,twocolumn,superscriptaddress,letterpaper]{revtex4}
\usepackage{amssymb}
\usepackage{amsmath}
\usepackage{graphicx}
\usepackage{natbib}
\usepackage{color}

\definecolor{darkblue}{rgb}{0,0,0.5}
\definecolor{lila}{rgb}{0.3,0,0.3}
\definecolor{turq}{rgb}{0,0.1,0.4}
\definecolor{lightblue}{rgb}{0.7,0.7,0.9}
\usepackage{url} % that hyperlinks may be hyphenated correctly
\usepackage[pdftex,
 colorlinks=true,
 backref=page,
 linkcolor=darkblue, % usual links
 filecolor=red,
 citecolor=turq, % for bibliographic
 urlcolor=lila, % for Emails etc
 pdftitle={Extinction of Light and Coherent Scattering by a Single Nitrogen-Vacancy Center in Diamond},
 pdfauthor={Thai Hien Tran and Jorg Wrachtrup and Ilja Gerhardt},
 pdfsubject={},
 pdfkeywords={},
 pdfpagelabels=true, % damit die Nummerierung stimmt
 breaklinks=false,
 plainpages=false,
 backref=false,
 bookmarks,
 bookmarksnumbered=true]{hyperref}

\begin{document}

\title{Extinction of Light and Coherent Scattering\\ by a Single Nitrogen-Vacancy Center in Diamond}

\author{Thai Hien Tran}
\affiliation{3. Institute of Physics, University of Stuttgart and Institute for Quantum Science and Technology, IQST, Pfaffenwaldring 57, D-70569 Stuttgart, Germany}
\author{J\"org Wrachtrup}
\affiliation{3. Institute of Physics, University of Stuttgart and Institute for Quantum Science and Technology, IQST, Pfaffenwaldring 57, D-70569 Stuttgart, Germany}
\affiliation{Max Planck Institute for Solid State Research, Heisenbergstra\ss e 1, D-70569 Stuttgart, Germany}
\author{Ilja Gerhardt}
\affiliation{3. Institute of Physics, University of Stuttgart and Institute for Quantum Science and Technology, IQST, Pfaffenwaldring 57, D-70569 Stuttgart, Germany}
\affiliation{Max Planck Institute for Solid State Research, Heisenbergstra\ss e 1, D-70569 Stuttgart, Germany}
\email{Corresponding author: i.gerhardt@fkf.mpg.de}

%TC:ignore
\begin{abstract}
Coherently scattered light from a single quantum system promises to get a valuable quantum resource. In this letter an external laser field is efficiently coupled to a single nitrogen vacancy \mbox{(NV-)}center in diamond. By this it is possible to detect a direct extinction signal and estimate the NV's extinction cross-section. The exact amount of coherent and incoherent photons is determined against the saturation parameter, and reveals the optimal point of generating coherently scattered photons and an optimal point of excitation. A theoretical model of spectral diffusion allows to explain the deviation to an atom in free-space. The introduced experimental techniques are used to determine the properties of the tight focusing in an interference experiment, and allow for a direct determination of the Gouy-phase in a strongly focused beam.
\end{abstract}
%TC:endignore
\maketitle

The negatively charged nitrogen vacancy (NV-) center in diamond has been a steady source for research in the past two decades. It allows for quantum sensing applications~\cite{balasubramanian_n_2008,staudacher_s_2013}, and a variety of quantum optical primitives. These cover the single-shot readout of a nuclear spin~\cite{neumann_s_2010}, quantum error correction~\cite{waldherr_n_2014}, or the coupling of neighboring defects~\cite{dolde_np_2013} to implement a basic quantum network. While the described experiments solely rely on an optical readout of the electron (and eventually accessible nuclear) spin, another line of experiments deals with the distant interaction of NV-centers by means of optical photons~\cite{sipahigil_prl_2012}. These experiments are conducted under cryogenic conditions, since they require the coherence of the photons to the internal spin state~\cite{yang_np_2016}. This is the prerequisite for spin photon entanglement. Unfortunately, also under cryogenic conditions, the amount of coherently emitted photons is limited. This is due to the dominating phonon side band in the NV's emission. Only about 3-4\% of the light resides on the zero-phonon-line (ZPL)~\cite{jelezko_pssa_2006}.

The theory of coherent scattering of a single emitter is well described since the 1960s~\cite{glauber_pr_1963,mollow_pr_1969}. Early experiments on single atoms or ions showed a typical behavior as expected from a two-level system~\cite{kimble_prl_1977,wineland_ol_1987}. The spectroscopy on solid-state emitters, such as molecules and quantum dots involve more effects on the amount of coherent scattering, such as phonon-contributions. Recent progress has shifted the focus in single emitter spectroscopy from a high \emph{collection efficiency} towards an \emph{efficient coupling} of an external field towards an emitter. In the past decade such measurements have been performed~\cite{gerhardt_prl_2007,wrigge_np_2008,vamivakas_nl_2007,tey_np_2008}, and allow for ultra-narrow-band photons~\cite{matthiesen_prl_2012}, and enable squeezing measurements on the light of single emitters~\cite{schulte_n_2015}.

\begin{figure}[h!b]
  \includegraphics[width=\columnwidth]{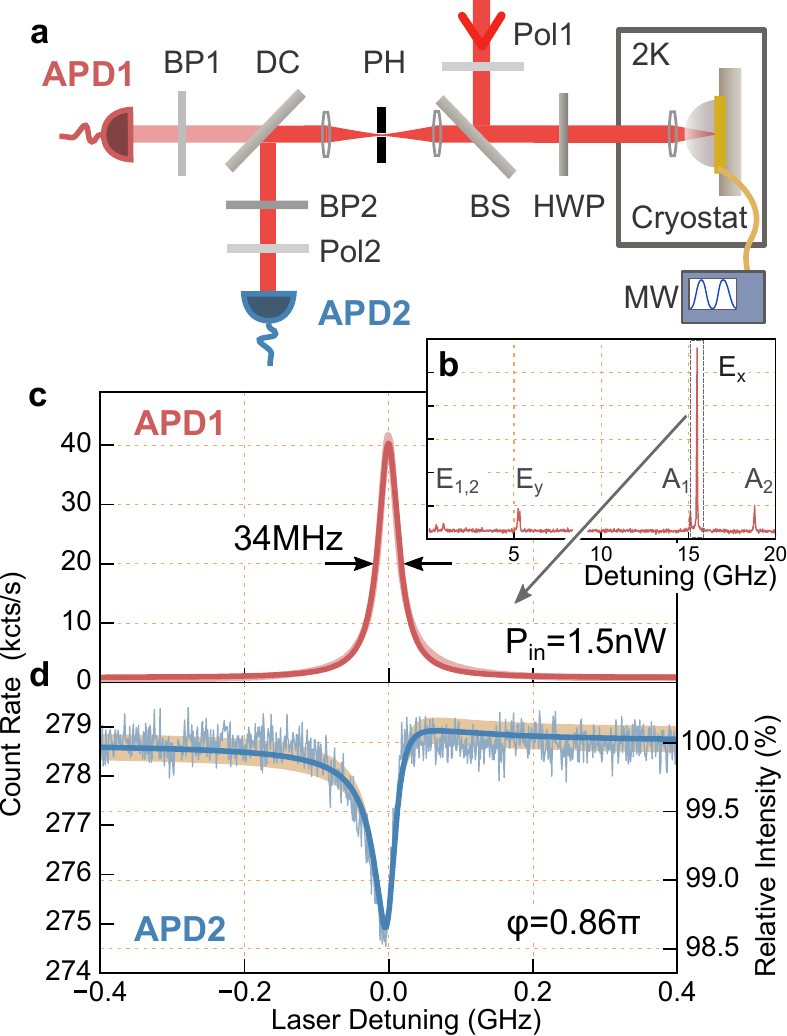}
  \caption{a) The experimental setup consists of a cryogenic confocal microscope. Two detection arms analyze the emission of the photon side band (APD1) and the coherent laser light (APD2) b) Fluorescence excitation spectrum behind a 650-750~nm band-pass filter (BP1)~\citep{batalov_prl_2009}. c) Zoom on the $E_{\mathrm{x}}$ transition. d) Coherent scattering of the NV- and its interference with the incoming laser. Pol: Polarizer; HWP: half-wave plate; PH: Pinhole; DC: Dichroic; MW: microwave.}
  \label{fig:fig01}
\end{figure}

Here we show a measurement of the coherent scattering of a single nitrogen-vacancy center. This is recorded by monitoring the direct extinction of light by the nanoscopic emitter. \emph{Direct} refers in this context to a simple observing of an altered laser with a single photon detector~\cite{plakhotnik_prl_2001}. As expected, the amount of coherent and incoherent scattered photons changes with the incident flux of photons~\cite{cc_book}. An interferometric application of this atom-sized defect is applied to a measurement of the Gouy-phase in the tight focus in our experimental configuration.

The extinction of light by a single emitter is one of the most evident proofs of coherent scattering~\cite{wineland_ol_1987,plakhotnik_prl_2001}. Usually, an incoming laser beam is altered by the presence of an emitter in the optical path. The amount of light, collected on a detector in the forward direction, is given as $I_{\mathrm{det}}=\langle(\mathbf{E}_{\mathrm{laser}}+\mathbf{E}_{\mathrm{emitter}})^2\rangle$, i.e.\ the incident laser light and the coherently scattered light interfere destructively in the optical far-field. This is commonly described as \emph{extinction}. In the case of a purely coherently scattering emitter, such as an atom at the low excitation limit, the light is simply reflected back to the exciting laser source~\cite{zumofen_prl_2008} and the light in the forward direction can be perfectly extinguished.

Extinction measurements can not only be performed in the forward direction~\cite{plakhotnik_prl_2001}. Generally, the interference occurs into all directions, but as outlined in the optical theorem~\cite{newton_ajop_1976} the extinction in the forward and the backward direction amounts to the same value as the \emph{absorption}, which is the transfer to other forms of energy, such as heat. If parts of the laser light are reflected elsewhere, or coupled deliberately onto the detector, the resulting interference signal can have an arbitrary phase, $\varphi$, depending on the relative phase of the reflected light to the coherent scattering of the emitter. For an emitter with a lifetime-limited line it is possible to rewrite the above equation and describe the signal on the detector as

\begin{equation}
I_{\mathrm{det}}=I_{\mathrm{laser}}\left(1-\mathcal{V}\frac{\varGamma_2(\Delta\cos{\varphi}+\varGamma_2\sin{\varphi})}{\Delta^2+\varGamma_{\mathrm{eff}}^2}\right) \, .
\label{eqn:eqn01}
\end{equation}

\noindent
with an reflected flux of the laser of $I_{\mathrm{laser}}$, the visibility $\mathcal{V}$, the homogeneous linewidth $2\varGamma_2$, the effective, e.g.\ power-broadened, linewidth $2\varGamma_{\mathrm{eff}}$, and a spectral detuning $\Delta$. The \emph{effective} visibility, or contrast, reduces with increasing excitation power expressed in the saturation parameter $S$ as

\begin{equation}
	\mathcal{C}(S)=\mathcal{V} \frac{1}{S+1} \, .
	\label{eqn:eqn02}
\end{equation}

\noindent
This measurable contrast, $\mathcal{C}$, is described by the fraction of missing light. It depends on the orientation of the emitter against the laser field, the amount of coherent vs.\ incoherent scattering, the amount of collected light from the emitter and the laser and other diminishing factors, such as the Debye-Waller-factor $\alpha_\mathrm{DW}$. In the presented case with a single NV-center, only a small fraction of the overall emission~\cite{jelezko_pssa_2006} leads to the interference due to the low spectral coherence of the emitted photons. This diminishes the effective extinction cross-section, $\sigma$, to a fraction of the maximal theoretical possible value of $3 \lambda^2 / (2 \pi)$, and extinction experiments with NV-centers are hard, and commonly require balanced detection or lock-in techniques~\cite{buckley_s_2010}.

Fig.~\ref{fig:fig01}a shows the experimental setup. A tunable narrow-band laser (New-Focus Velocity $\approx$ 637~nm) excites a single NV-center inside a bath cryostat (T=2~K). Tight focusing is realized by a microscope objective (Zeiss A-Plan, 40$\times$, 0.65NA) in conjunction with a hemispherical macroscopic ($\varnothing$=1~mm) monolithic diamond solid-immersion lens (SIL, Element Six). It exhibits a 0.6~mm thick overgrown layer of ultra-pure diamond on the flat side. This contains single natural NV-centers~\cite{siyushev_apl_2010} suitable for cryogenic experiments. The polarization of the incident laser is laterally aligned to the $E_{\mathrm{x}}$-transition (Fig.~\ref{fig:fig01}b).

Due to strain-induced mixing of the excited-state spin sub-levels, optical excitation might lead to a spin flip. Since the optical transitions are spin-selective, these reduce the excitation efficiency. This is avoided by applying a microwave field ($\nu$=2.87~GHz) to keep the desired spin-state always accessible. Resonant excitation might induce a 2-photon-process, that leads to an ionization of the NV-center into the neutral charge state~\cite{siyushev_prl_2013}. If no fluorescence is observed in a frequency scan, a 300~ms green, 20~$\mu$W 532~nm laser pulse is applied until the fluorescence is observed again.

The emission of the phonon side band (PSB) is captured on a single photon detector (APD1) behind a band-pass filter (650-750~nm). The laser back-scattering and the zero-phonon line (ZPL) is captured on another single photon detector (APD2). This light is reflected off a dichroic (Chroma ZT640) and passes a narrow-band filter (637$\pm$1~nm, Omega Optical). The polarizer (Pol2) is aligned to the incident laser field. The count rate of both detectors is monitored against laser detuning of the excitation laser.

Fig.~\ref{fig:fig01}b shows the red-shifted emission from the NV- against the laser detuning. A finer resolved spectrum is displayed in Fig.~\ref{fig:fig01}c. It is assembled of 2000 single spectra (each consisting of 1000 frequency pixels a 2~msec), shifted according to their spectral center to suppress spectral diffusion. This is determined by fitting a Lorentzian line to each of the lines. The resulting spectral linewidth (34~MHz) at 1.5~nW laser excitation (measured in front of the cryostat) shows that we exceed the lifetime limited linewidth (full width at half maximum, FWHM) of $\varGamma_1=1/(2 \pi T_1)\approx$13~MHz.

Simultaneously, the extinction of light is monitored on APD2 (Fig.~\ref{fig:fig01}d). Far off-resonance, the signal is given by laser reflection mainly from the flat side of the SIL. Power fluctuations are suppressed to the shot-noise limit by a PID controller (SRS, PID960) in the excitation arm. When the laser is detuned and the emitter is on resonance, an extinction signal is observed as described by Eqn.~\ref{eqn:eqn01}. Note that the detection pinhole (50~$\mu$m) is aligned such that the emission of the NV-center is fully captured, whereas the laser reflection is (partially) out of focus and therefore suppressed by three orders of magnitude.

The final phase $\varphi$ of 0.86$\pi$ to the exciting laser results from the depth of the emitter $d$, and the refractive index $n$ as $\varphi_{\mathrm{d}} =\pi - 2\pi ~(2dn \bmod \lambda)$ and the phase difference of the net laser reflection on all interfaces. If no surfaces and no laser reflection would be present, a simple Lorentzian reflection of the coherent scattering would be observed~\cite{zumofen_prl_2008}. The dominant noise source is the photon shot noise (orange band in Fig.~\ref{fig:fig01}c, $\pm$1~$\sigma$). The spectra are also corrected from an internal cavity, which results from the cryostat windows. In summary, a signal to noise ratio of 15 is observed for this measurement. These experiments were also performed at the low excitation limit (not shown). Correspondingly, the measured contrast is increased as $1/(S+1)$, with $S$ being the saturation parameter on resonance $\Omega^2/(\varGamma_1\varGamma_2)$~\cite{cc_book} which depends on the excitation power $\Omega^2$. Subsequently, we measure a maximal contrast $\mathcal{C}$ of 2.8\% at the lower excitation limit. At this point of excitation this value corresponds to the visibility $\mathcal{V}$.

\begin{figure}[ht]
  \includegraphics[width=\columnwidth]{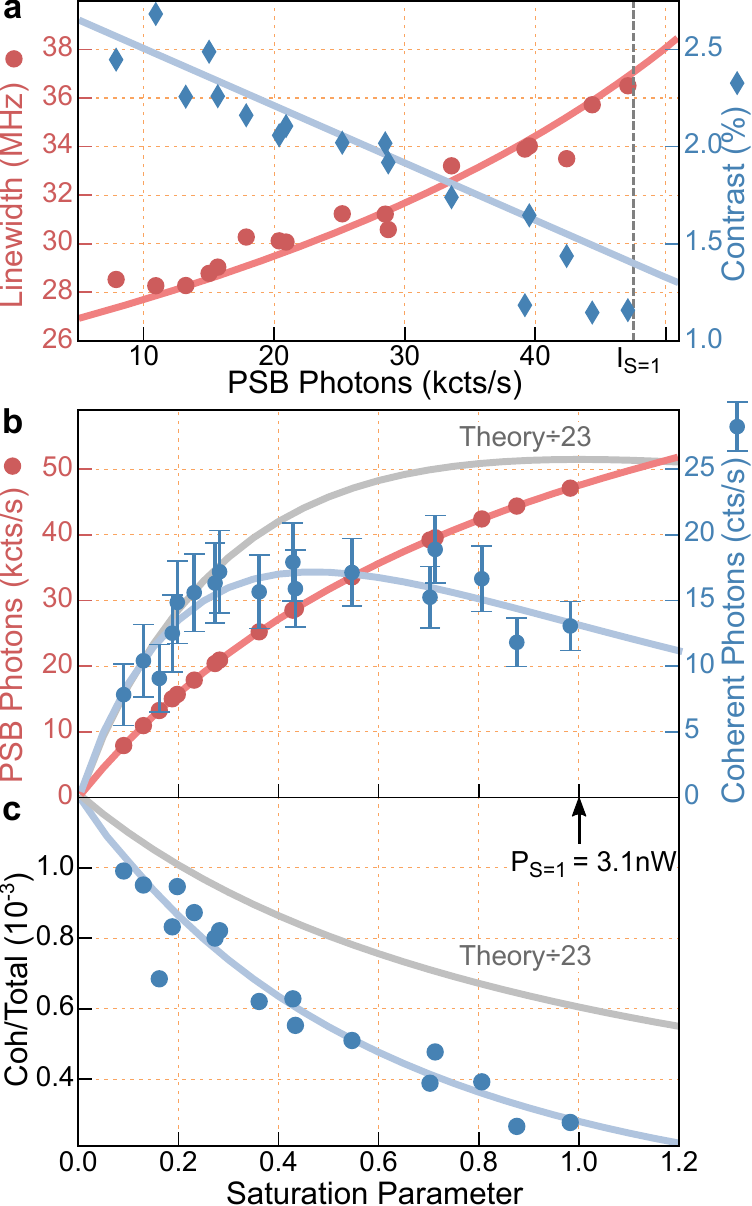}
  \caption{a) Determination of the saturation parameter by the linewidth measured on APD1 (red). Contrast, measured on APD2, fitted by Eqn.~\ref{eqn:eqn01}. The dashed line indicates saturation intensity $I_{\mathrm{sat}}$ b) Amount of coherent (blue line) and PSB (red line) photons, as deduced by the data in a). c) Ratio of coherent photons to total emission. b,c) Gray curves: theoretical amount and ratio of coherent photons calculated from the total collected PSB emission.}
  \label{fig:fig02}
\end{figure}

The observed signal on APD2 consists not only of the laser reflected from the interfaces and the coherent scattering of the NV-center. It might also contain a small amount of red-shifted fluorescence, which is present when the NV-center is on resonance. This can not interfere with the incident laser light. Therefore, all extinction measurements such as Fig.~\ref{fig:fig01}d contain both contributions. A full procedure to derive both components independently with the use of a polarizer~\cite{wrigge_np_2008} is described in the supplementary material. Note that under close to crossed polarizer conditions, the contrast rises well above 5\%.

As the extinction signal, the amount of coherent and incoherent scattering is power dependent with the saturation parameter $S$. This unit-less entity also influences the linewidth in the red-shifted fluorescence of the NV-center on APD1 (Fig.~\ref{fig:fig02}a, red dots). The curve (red line) is fitted as $\varGamma_{\mathrm{eff}} = \varGamma_2 ~ \sqrt{\frac{I_{\mathrm{sat}}}{I_{\mathrm{sat}}-I}}$, and allows to determine $I_{\mathrm{sat}}$ and $\varGamma_2$. $I_{\mathrm{sat}}$ then reveals the saturation parameter $S=\frac{I}{I_{\mathrm{sat}}-I}$ for the following measurements. 

Fig.~\ref{fig:fig02}a (red) represents one measurement as in Fig.~\ref{fig:fig01}c. The excitation power ranges from 0.3~nW to 3.0~nW, corresponding to a saturation parameter up to 1.0. Due to the low signal-to-noise ratio of the single scans at very low incident powers ($P_{\mathrm{in}}<$ nW, APD2$<$ 1.5~kcts/s), the fit error is increased and leads to an uncertainty in the fitted linewidth. The same holds for very high excitation powers, where the NV's spectral diffusion and the probability of ionization is increased. Since the measurement expands over several cooling cycles, cavity and thermal effects change the coupling and collection efficiency slightly. Therefore, we decided to use the PSB intensity as an indirect reference for the saturation parameter instead of the incoming power. The relationship between the laser power and PSB intensity is independently validated by the recording of a saturation curve.

The contrast of the extinction signal is simultaneously recorded on APD2 (Fig.~\ref{fig:fig02}a, blue). This depends on the amount of coherently scattered photons. With an increasing excitation power it is reduced as $1/(S+1)$. A full derivation of this procedure, and its independence to the incoherent amount, leaking through the narrow-band filter, is derived in the supplementary material. 

The experimental determined amount of coherent scattering by the single NV-center is displayed in Fig.~\ref{fig:fig02}b, blue. For a two level system, with a collection efficiency of $\eta$ the amount of coherent scattering is expected to obey the equation $I_{\mathrm{coh}}=\eta \alpha_{\mathrm{DW}} \frac{\varGamma_1^2}{4\varGamma_2} \frac{S}{(1+S)^2}$ (Fig.~\ref{fig:fig02}b, gray). It implies that the amount of coherent scattering is highest at a saturation parameter of unity. Instead, we find the amount of coherent scattering reduced against its predicted value. This is expected since residual charge noise in the environment with the NV-center increases exponentially with the excitation power as $\exp(-P_{\mathrm{in}}/P_{\mathrm{env}})$. With this model (solid curve in Fig.~\ref{fig:fig02}b), assuming $P_{\mathrm{env}}$= 4.0~nW, the amount of coherent photons is described well. Simultaneously, the incoherent scattering is monitored (red).

The optimal point to extract coherent photons corresponds to a saturation parameter of 0.45. In our experimental configuration this corresponds to 1.5~nW in front of the cryostat. This is the point where the largest ratio of coherent scattering vs.\ incident light intensity and also the most coherent scattering is observed. This point should be determined for all experiments which utilize coherent photons. Fig.~\ref{fig:fig02}c shows the ratio of coherent to incoherent scattering, while the description above gives the absolute range. Theoretically this obeys the factor $1/(1+S)$, but is modified here due to charge noise from the environment as $\exp(-P_{\mathrm{in}}/P_{\mathrm{env}})/(1+S)$.

The extinction cross-section $\sigma_{\mathrm{ext}}$ influences the contrast as $\sigma_{\mathrm{ext}}/A$ with the focal area $A$. The full set of experiments is used to deduce the extinction cross-section of the NV-center. It depends on various factors, such as the Debye-Waller factor (3-4\%), and the NV's physical orientation. The latter is deduced from the orientation of the diamond lattice and the incident polarization of the incoming laser. Together with the achievable extinction signal and its saturation behavior, we estimate the extinction cross-section to $\varnothing$ 30~nm. This value is now estimated for an experiment in the forward direction, such that the entire light beam would be captured by a single-photon detector. This value is strongly polarization dependent as $\cos{(\theta)}^2$ with $\theta$ as the relative angle between the NV-axis (with $E_{\mathrm{x}}$ being a linear dipole) and the laser. Therefore, it will be possible to influence an ongoing laser beam with the NV's spin state, e.g.\ on the $A_1$ or the $A_2$ transition. The achievable effect is in the order of 0.23\%. Of course, the measurable contrast can be artificially enhanced, by neglecting photons in the forward direction, e.g.\ in a crossed polarization configuration~\cite{gerhardt_prl_2007,buckley_s_2010}.

Since the scattered photons do not necessarily lead to an excitation of the NV-center, many photons may be scattered at the low excitation limit before the system undergoes an excitation and a projective measurement. Therefore many quantum optical primitives, such as quantum non-demolition measurements can be realized with this effect. Another option are interferometric applications: The Gouy phase, $\phi(z)$ describes the ``inversion'' of the wave-fronts in an optical focus~\cite{gouy_1890,feng_ol_2001}.

\begin{figure}[ht]
  \includegraphics[width=\columnwidth]{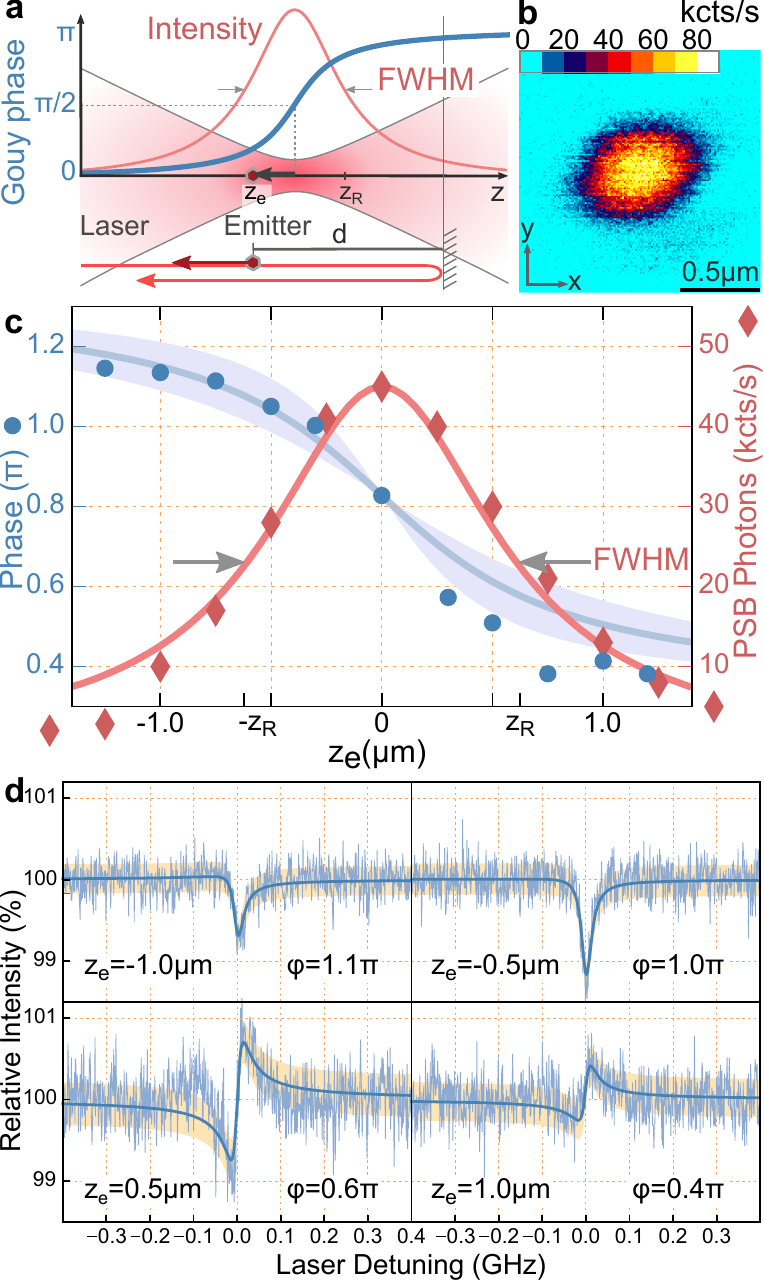}
  \caption{Determination of the Gouy-Phase, utilizing the NV's E$_x$ transition. a) Optical focus (gray line), intensity distribution (red), and the Gouy-phase (blue). b) The lateral scan at the focus, $w_0$=230~nm. c) The phase of the interference signal depends on the displacement of the emitter $z_e$. The phase change represents the Gouy phase. d) Samples of the signal at different axial positions, $z_e$.}
  \label{fig:fig03}
\end{figure}

When the emitter is placed in the focused laser (Fig.~\ref{fig:fig03}a), where $z_{\mathrm{e}}$ is its axial position relative to the laser focus, the Gouy phase is observed by the resulting change of the phase of the interferometric signal~\cite{hwang_oc_2007}. The phase difference between $z_{\mathrm{e}}=-\infty$ and $z_{\mathrm{e}}=+\infty$ is the total Gouy phase shift $\pi$, and its highest gradient is at the focus.

To measure the Gouy-phase, the NV-center is axially moved through the focus of the confocal microscope. By an earlier lateral scan (See Fig.~\ref{fig:fig03}b), a focus waist $w_0$ of 0.23$\pm$0.05~$\mathrm{\mu m}$ was estimated. This corresponds to a Rayleigh length $z_{\mathrm{R}}$ of $0.62\substack{+0.30 \\ -0.24}~\mathrm{\mu m}$. The Gouy-phase is measured in a range of $\pm$2.5~$z_{\mathrm{R}}$, corresponding in an expected change of $\Delta\varphi \approx $0.8$\pi$. The excitation power is adjusted, that far in front and behind the focus the count rates are not dramatically reduced. The intensity distribution (i.e.\ power per area) is monitored along with the interferometric detection. The count rates, measured on APD1, normalized to the excitation intensity in the focus, are depicted in Fig.~\ref{fig:fig03}c, red.

Simultaneously, the extinction of light is monitored on APD2. The phase $\varphi$ reveals directly the information on the Gouy-phase. Sample measurements are shown in Fig.~\ref{fig:fig03}c (Fig.~\ref{fig:fig01} is in focus). The only fit parameters being the residual phase offset (0.86$\pi$ in the focus) and the collection efficiency, since all other parameters ($z_\mathrm{R}$, $w_0$) are determined independently. The blue band in the back indicates the error bar for estimating the Rayleigh length, $z_\mathrm{R}$, as outlined above.

In conclusion, the presented measurements prove the ability to influence a laser field with an atomic sized single quantum emitter and to reach up to 2.8\% direct influence on a back reflected laser. The achievable effect is much higher when the polarizations of the excitation and detection are not aligned as in the presented experiment, and more laser light is suppressed. For the forward direction, we calculate an achievable effect of 0.42\%. This implies that approx.\ every 250th photon interacts with the emitter. Comparable experiments will allow to implement a variety of quantum optical primitives, such as a spin-dependent phase gate. Another line are quantum non-demolition experiments, which hold the promise of producing larger cluster states of light~\cite{rao_prb_2015} or implement other schemes in quantum computing~\cite{munro_njp_2005}. In the context of precision measurements, we like to underline that the described visibility is also accessible under extremely low incident flux. For the low excitation limit this can exceed the signal to noise ratio of fluorescence detection~\cite{wrigge_np_2008,wrigge_oe_2008}. A first measurement reveals an interferometric measurement of the Gouy phase with an NV-center. This is comparable to earlier single ion experiments in the gas phase~\cite{hetet_prl_2011}, but is now extended to strong focusing in the solid state.

\begin{acknowledgments}
We thank Dr.\ P.\ Siyushev, who conducted an initial experiment. We acknowledge the funding from the MPG, the SFB project CO.CO.MAT/TR21, the BMBF, the project Q.COM, and SQUTEC.
\end{acknowledgments}

\end{document}